\documentclass[useAMS,usenatbib]{mn2e}
\usepackage{graphicx}
\usepackage{txfonts}
\usepackage{natbib}

\def\m{\rm M82\thinspace X-1}
\def\suzaku{{\it SUZAKU}}
\def\heasoftv{\hbox{\rm{\small HEASOFT~\/}}}
\def\xis{{\rm XIS}}
\def\xselect{\hbox{\rm{\small XSELECT~\/}}}
\def\addascaspec{\hbox{\rm{\small ADDASCASPEC~\/}}}
\def\ftool{\hbox{\rm{\small FTOOL}}}
\def\grppha{\hbox{\rm{\small GRPPHA~\/}}}
\def\kev{\hbox{$\rm\thinspace keV$}}
\def\hxd{{\rm HXD}}
\def\hxddtcor{\hbox{\rm{\small HXDDTCOR}}}
\def\ctps{\hbox{$\rm\thinspace count~s^{-1}$}}
\def\pin{{\rm PIN}}
\def\ks{\hbox{$\rm\thinspace ks$}}
\def\mgtime{\hbox{\rm{\small MGTIME}}}
\def\mathpha{\hbox{\rm{\small MATHPHA}}}
\def\chisq{{\chi^{2}}}

\title[]{Weakly broadened iron line in the X-ray spectrum of the Ultra Luminous X-ray source {\rm M82\thinspace X-1}}

\author[M.~D. Caballero-Garc\'{i}a]{M. D. Caballero-Garc\'{i}a$^{1,}$$^{2}$\thanks{E-mail:
mcaballe@ast.cam.ac.uk} \\
$^{1}$Institute of Astronomy, Madingley Road, Cambridge, CB3 0HA, UK \\
$^{2}$Department of Physics, University of Crete, GR-71003, Heraklion, Greece
}

\begin{document}


\pagerange{\pageref{firstpage}--\pageref{lastpage}} \pubyear{2002}

\maketitle

\label{firstpage}

\begin{abstract}
In this paper we present the best quality {\it XMM-Newton} and {\it Suzaku} data from \m so far. We analyze the spectra 
of this remarkable Ultra-Luminous X-ray Source in a self-consistent manner. We have disentangled emission from the host
galaxy, responsible for the soft X-ray emission ($E{\le}2.5$\,keV), which is successfully described by a two-temperature 
thermal emission from a hot plasma in multi-phase state, plus a narrow Fe line emission at 6.7\,keV. This allowed us to 
properly study the intrinsic continuum emission from \m. The continuum of the {\it Suzaku} spectrum is curved 
and the high quality data of the {\it Suzaku} spectrum allowed us to significantly detect a weakly broadened
Fe K${\alpha}$ emission line. The Equivalent-Width is in the range 30--80\,eV and it does not depend on the model 
applied for the continuum. Assuming that this line is coming from the ULX via disc fluorescence, the data indicates 
a disc truncation at a radius of 6--20 gravitational radii. This value is comparable to or even larger than the Innermost 
Stable Circular Orbit of a non-spinning (Schwarzschild) black hole. Future longer observations might test this scenario.
\end{abstract}


\section{Introduction}

Ultra-luminous X-ray (ULX) sources are point like, off-nuclear extra Galactic sources with observed luminosities 
greater than the Eddington limit for a $10\,{\rm M}_{\odot}$ black hole (BH), with $L_{X}{\ge}10^{39}$\,${\rm erg\,s^{-1}}$ \citep{f1}. 
The true nature of these objects is still open to debate \citep{mc1}. One fundamental issue is whether the emission 
is isotropic or beamed along our line-of-sight. A possible scenario for geometrical beaming involves super-Eddington 
accretion during phases of thermal-timescale mass transfer \citep{k1}. Alternatively, if the emission is isotropic and 
the Eddington limit is not violated, ULX must be fuelled by accretion onto Intermediate-Mass BH (IMBH), with masses in 
the range 100-10\,000 ${\rm M}_{\odot}$ \citep{cm1}. Of course, it is possible that some ULX appear very luminous due to 
a combination of moderately high mass, mild beaming and mild super-Eddington emission. It is also possible that ULX are an 
inhomogeneous population, comprised of both a subsample of IMBH and moderately beamed stellar mass black holes \citep{f2,mc1}.

\m\ (also named CXOU J095550.2+694047) is one of the brightest ULX in the sky. The host galaxy is located nearby, at a distance 
of 3.6\,Mpc \citep{freedman94}, making \m, with an average X-ray luminosity ${\gtrsim}10^{40}$\,${\rm erg\,s^{-1}}$, an excellent 
target for the detailed modelling of the X-ray spectrum in such objects. \m\ is one of the few ULX for which (10-200\,mHz) 
Quasi-Periodic Oscillations (QPO) have been found \citep{stroh03} and is the prototype ULX with curved X-ray spectra. 
\citet{miyawaki09} analyzed the {\it Suzaku} spectrum of this source (with an energy coverage extending up to to 20\,keV) and 
tested the data to a variety of models, confirming that this source has a spectrum curved at high-energies. \citet{mucciarelli06} 
analyzed the longest {\it XMM-Newton} observations available in the archive and applied models of diffuse thermal emission from 
the galaxy to understand the high-energy curvature of the spectrum. The authors found that the high energy curvature could indeed  
be explained by the presence of a third thermal component having very high column densities ($N_{H}{\approx}10^{23}$\,${\rm cm^{-2}}$).
This contradicts recent {\it CHANDRA} observations by \citet{feng10}, who found column densities of 
$N_{H}{\approx}1{\times}10^{22}$\,${\rm cm^{-2}}$, in fitting the X-ray spectra at the position of this ULX.

As in the case of black hole binaries (BHBs), some ULXs undergo spectral transitions from a power-law dominated state to a ``high-soft'' 
state (see e.g. McClintock \& Remillard for a description of the spectral states in BHBs). During the power-law state
the high-energy spectra of ULXs is inferred to have a power-law spectral shape in the 3-8\,keV spectral range, together with a high-energy 
turn-over, and a soft-excess (e.g. \citealt{kaaret06}). This soft-energy excess can be modelled by emission coming from the inner-accretion 
disc and is characterized by a low disc temperature of ${\approx}0.2$\,keV (\citealt{miller03,miller04}). If these observed characteristics 
are entirely due to a BH accreting at sub-Eddington rates, they imply the presence of an IMBH with $M{\approx}100-1000$\,${\rm M}_{\odot}$. 
In the case of \m, due to the huge absorption at soft X-rays by the host galaxy, this soft X-ray excess is not observed in the spectra. 
Systematic studies of the highest quality XMM-Newton ULX spectra by \citet{yoshida10} and \citet{stobbart06} have shown that the majority 
of ULXs display a break/turnover at high energies (${\gtrsim}3$\, keV). Such breaks are not commonly seen in the spectra of BHBs. 
\citet{gladstone09} argue that this turnover could be due to Comptonization from an optically thick corona, which would shroud the inner 
regions of the accretion disc and artificially lower the inner temperatures obtained from simple spectral modelling. Under such interpretation 
most ULXs would then represent a new accretion state for stellar-mass BHBs. An alternative explanation is proposed by \citet{caba10}, where 
the authors note that in many cases the turnover occurs between 5--7 keV, and could be due to a combination of the Iron ${\rm K}_{\alpha}$ 
emission line and absorption edge (6.4 and 7.1 keV respectively for neutral Iron) in a relativistically blurred reflection spectrum from
the inner accretion disc of a spinning black hole. Here reflection refers to the  back-scattering and fluorescence of X-rays \citep{george91}.
Under this interpretation, the soft excess emission often assumed to come directly from the accretion disc is also due to blurred reflection, 
meaning that the disc emission is not directly observed. Other explanations have also been proposed, often based on emission from the 
aforementioned slim discs in which advection of radiation from the inner regions of the accretion disc reduces the observed luminosity 
(Abramowicz et al. 1988). In the case of \m, the huge intrinsic column density of the galaxy ($N_{H}{\approx}10^{22}$\,${\rm cm^{-2}}$), 
makes the detection of soft X-ray spectral features from this source very unlikely. 

Here we use the best publicly available data from {\it Suzaku} and {\it XMM-Newton}, currently being the best available X-ray 
data from this source. Our goal is to disentangle both the emission from the host galaxy and the intrinsic spectrum from the source (Section
\ref{spec_anal}) and to analyze the latter with current models used for the description of the spectra from ULXs (Section \ref{spec_anal2}). We notice the presence 
of a broad Fe K line in the residuals and explore its significance in Section \ref{spec_anal3}. Eventually, physical implications are discussed in Section \ref{discuss}.

\section{Observations and data reduction} \label{observ}

\subsection{{\it XMM-Newton}}

In this work, we consider the longest publicly available {\it XMM-Newton} EPIC-pn spectra from \m\ (Obs. IDs: 0112290201 and 0206080101). 
The source was very bright, with a count rate of $3.06{\pm}0.012, 2.341{\pm}0.007\,cts/s$. The datasets were obtained through the {\it XMM-Newton} 
public data archive. The EPIC-pn camera has a higher effective area than the EPIC-MOS cameras, and drives the results
of any joint spectral analysis. The reduction and analysis reported in this work used SAS version 9.0.0. We checked for pile-up in all
the observations and found that this was not significant (i.e. it is ${\le}5\%$ for the high energy channels) for both observations.

In Table \ref{log_obs} we present a log of the observations. We applied the standard time and flare filtering (rejecting high-background 
periods of rate ${\ge}0.4$\,counts/s, as recommended for the pn camera \footnote{Information provided at "node52.html" of the User Scientific Guide.}). 
We filtered the event files, selecting only the best-calibrated events (pattern${\le}4$ for the pn), and rejecting flagged events (flag$=0$).

For each exposure, we extracted the flux from a circular region on the source (centre at coordinates ${\rm RA}=9\,h55\,m50.2\,s$\,deg, 
${\rm Dec}=69\,d40\,m47\,s$\,deg and radius $18\,$arcsec). The background was extracted from a circular region, not far from the 
source and away from boundaries of the chips and the nucleus of the galaxy. We built response functions with the SAS tasks {\tt rmfgen} and 
{\tt arfgen}. We fitted the background-subtracted spectra with standard models 
in XSPEC 12.5.0 \citep{a1}. All errors quoted in this work are $90\%$ confidence errors. For the spectral fitting we used the 0.5--10\,keV range. 
The resulting spectra were grouped with the FTOOL {\tt grppha} to bins with a minimum of 20 counts each.

\subsection{{\it Suzaku}}   \label{ins_suzaku}

\m\ was observed with \suzaku\ (\citealt{SUZAKU}) on three occasions in 2005 October 4, 19 and 27 for approximately 32, 40 and 28\ks\ respectively. 
The four operating detectors constituting the X-ray Imaging Spectrometer (XIS; Koyama et al. 2007) were  operated in the ``normal'' clock mode in 
all observations. In all cases, the detectors were operated in both the 3x3 and 5x5 editing modes.  A total (co-added) front-illuminated (FI) exposure 
of approximately 96, 110 and 80\ks\ were obtained for the three respective observations. The corresponding back-illuminated (BI) exposures were 
approximately 32, 37 and 27\ks. Using the latest \heasoftv software package we processed the unfiltered event files for each of the \xis\ CCDs and editing modes
operational in the respective observations, following the \suzaku\ Data Reduction Guide\footnote{http://heasarc.gsfc.nasa.gov/docs/suzaku/analysis/}. 
We started by creating new cleaned event files by re-running the \suzaku\ pipeline with the latest calibration, as well as the associated screening criteria files.
\xselect was used to extract spectral products from these event files. In all observations, source  events were extracted from a circular region of 
180'' radius centred on the point source, and background spectra from another region of the same size, devoid of any obvious contaminating emission, 
elsewhere on the same chip. The script ``xisresp''{\footnote  {http://suzaku.gsfc.nasa.gov/docs/suzaku/analysis/xisresp}} with the ``medium'' input 
was used to obtain individual ancillary response files (arfs) and redistribution matrix files (rmfs). ``xisresp'' calls the tools ``xisrmfgen'' 
and ``xissimarfgen''. Finally, we combined the spectra and response files from the three front-illuminated instruments (XIS0, 2 and XIS3) using 
the \ftool\ \addascaspec. This procedure was repeated for each observation resulting in a total of six XIS spectra. After inspecting that there were 
no significant spectral differences between the three observations we further co-added the various front and back-illuminated spectra. This resulted in 
a total good exposure of approximately 280 and 92\ks\ for the front and back-illuminated spectrum respectively. Finally, the \ftool\ \grppha\ was used 
to give at least 20 counts per spectral bin. The two \xis\ spectra was fit in the 1.5--10.0\kev\  energy range with the 1.7--2.1\kev\ energy range being 
ignored due to the possible presence of un-modeled instrumental features.

For the hard X-ray detector (\hxd, \citealt{SUZ_HXD}) we again reprocessed the unfiltered event files for the
respective observations following the data reduction guide (only the \pin\ data is used in this analyses). Since the \hxd\ is a collimating
rather than an imaging instrument, estimating the background requires individual
consideration of the non X-ray instrumental background (NXB) and cosmic X-ray background
(CXB). The appropriate response and NXB files were downloaded for the respective
observations\footnote{http://www.astro.isas.ac.jp/suzaku/analysis/hxd/}; in each case the
tuned (Model D) background was used. Common good time intervals were obtained with
\mgtime\ which combines the good times of the event and background files, and \xselect\ was
used to extract spectral products. Dead time corrections were applied with \hxddtcor,
and the exposures of the NXB spectra were increased by a factor of ten, as instructed by the
data reduction guide. The contribution from the CXB was simulated using the form of
\cite{Boldt87}, with the appropriate normalisation for the nominal pointing (all observations
were performed with \hxd\ nominal pointing), resulting in a CXB rate of $\simeq 2.97\times10^{-2}$\ctps.
The NXB and CXB spectra were then combined using \mathpha\ to give a total background spectrum,
to which a 2\% systematic uncertainty was added. Similarly to the \xis\ data, we co-added all three \pin\ observations 
and grouped the final spectrum to have a minimum of 500 counts per energy bin to improve statistics, and again allow the use 
of $\chisq$ minimization during spectral fitting. The \pin\ data reduction yielded a total (co-added) source rate 
of $(1.5 \pm 0.3)\times10^{-2}$\ctps\ which is 2.5\% of the total observed flux, with a good exposure time of $\approx93$\ks.

\begin{table}
 \centering
 \begin{minipage}{120mm}
  \caption{{\it XMM-Newton} and {\it Suzaku} observations log.}
  \label{log_obs}
  \begin{tabular}{@{}lcccc@{}}
  \hline
   Satellite       &   Obs           &         Obs ID          & Date              & Exposure time (\,s)      \\
 \hline
   {\it} XMM-Newton   &   1          &    0112290201      &  2001-05-06     &   30\,558        \\
   {\it} XMM-Newton   &   2          &    0206080101      &  2004-04-21     &   104\,353       \\
   {\it Suzaku}       &   3          &    100033010       &  2005-10-04     &   32\,327        \\
   {\it Suzaku}       &   3          &    100033020       &  2005-10-19     &   40\,358        \\
   {\it Suzaku}       &   3          &    100033030       &  2005-10-27     &   28\,363        \\
\hline
\end{tabular}
\end{minipage}
\end{table}

\section{The determination of the emission from the diffuse component} \label{spec_anal}

The soft energies in the spectra of \m\ are dominated by the presence of X-ray emission lines having constant energies and width, thus indicating
an origin in the diffuse gas of the galaxy, rather than intrinsic to \m. We performed simultaneous fits of all the observations, considering constant 
properties (in time) for both thermal components of the diffuse emission from the galaxy (i.e. temperatures and element abundances tied between the 
observations, although see below for the latter). This is as expected, since the physical properties of the galaxy are likely to be constant with time.
To account for these lines we initially applied a single thermal plasma model with variable metal abundances ({\tt vmekal} model in XSPEC) to describe 
the softest part of the spectrum ($\le 2.5$\,keV). However, strong residuals were left with this model and we thus added a further second thermal component 
(as previously done in \citealt{mucciarelli06}) resulting in a significant improvement to the fit, with  the residuals at $\le 2.5$\,keV being almost null, 
within the instrumental errors. The values obtained for the spectral parameters are shown in Table \ref{table_phenom0}.
The value for the temperature of the hottest component in Obs. 2 found by \citet{mucciarelli06} ($kT=1.67{\pm}0.11$\,keV), 
is different possibly due to the addition of a further thermal disc component in their analyses. We tested for the presence of this further component and 
did not find it to be significant in the spectra.

We fixed metal abundances to solar values for Obs. 1 and 2 ({\it XMM-Newton} spectra), in agreement with previous work on {\it XMM}/{\it RGS} data \citep{read02}. 
Setting solar metal abundances for Obs. 3 ({\it Suzaku} spectrum) provided a very poor fit (${\chi}^{2}/{\nu}=1.92$, ${\nu}=3450$) and visible low energy residuals.
The best fit was achieved allowing metal abundances free to vary in the {\it Suzaku} spectra and they turned out to be super-solar for certain elements 
(${\rm Z}=6-8{\times}{\rm Z}_{\odot}$ for {\tt Si}, {\tt S} and {\tt Fe}). This fit apparently indicates super-solar abundances for the diffuse emission and therefore 
it is hard to reconcile with previous studies of metallicity of this galaxy (\citealt{origlia04,ranalli08}). These lines might have a different origin apart from 
the emission from the relatively {\it cold} emission from the galaxy (nevertheless, we refer to this emission as {\it hot} hereafter), which the model can not account for. 
Further studies of these emission lines require the use of additional emission processes/components that we will not consider in the present paper.

There is also a narrow line (i.e. ${\sigma}=0$) centred at ${\approx}6.7$\,keV coincident in energy with Fe ${\rm He}_{\alpha}$. We included it in all the fits
reported above and in the following fits as well. This line can not be accounted for the two thermal models (with solar abundances) employed to model the 
soft X-ray emission. We extracted spectra from other regions of the galaxy and confirm the presence of the narrow line. 
\citet{strickland07} spectrally resolved the Fe line at 6.7 keV with {\it Chandra} and {\it XMM-Newton} data and showed that it is primarily distributed diffusely 
rather than associated with discrete sources.

\begin{figure}
\centering
 \includegraphics[bb=34 18 553 744,width=5.5cm,angle=270,clip]{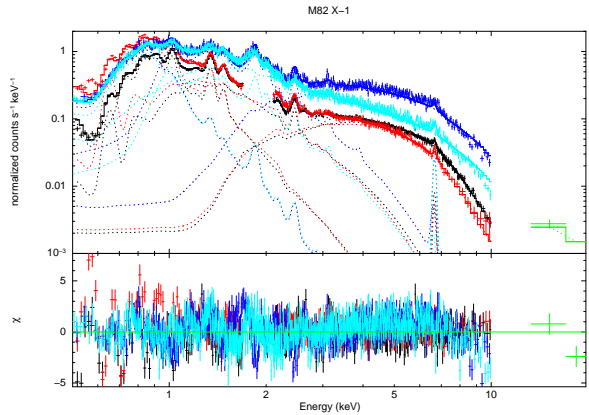}
 \caption{Global-Simultaneous fit of spectra of Obs. 1, 2 and 3 with a continuum phenomenological model ({\tt Highecut Powerlaw} for the high-energy emission). Diffuse emission from two hot plasmas (${\rm kT}=0.6$\,keV and ${\rm kT}=0.96$\,keV) and solar metallicities for {\it XMM-Newton} spectra and emission from a narrow Fe line have been taken into account. Black, red, blue and green correspond to FI,BI Suzaku/XIS, XMM-Newton and Suzaku/PIN/HXD spectra, respectively.}
 \label{plots_phenom}
\end{figure}

Having confirmed that the combination of two thermal plasma components plus emission from a narrow Fe line indeed provides a first approximation description of the 
emission from the diffuse component in the global-simultaneous fits, we proceed with this combination in all fits presented hereafter. As previously noticed by 
\citet{miyawaki09} for Obs. 3 (for which the energy coverage is maximum $0.5-20$\,keV in 
contrast to $0.5-10$\,keV for the remainder observations), the high-energy spectrum is strongly curved. We applied a curved phenomenological model to describe the 
high-energy spectrum ({\tt high-energy cut-off} in XSPEC). From the global simultaneous fits, the cut-off for Obs. 3 was constrained to be ${\rm E}_{\rm c}=6.83{\pm}0.19$\,keV. 
\citet{mizuno07} reported a very similar cut-off at 6-7\,keV in the spectrum of NGC~1313~X--1. The results of the global-simultaneous fits are shown in Figure 
\ref{plots_phenom} and Table \ref{table_phenom0}.

\begin{table}
 \centering
  \caption{Best-fit model spectral parameters with $90\%$ confidence errors obtained for the Global-Simultaneous fit shown in Figure \ref{plots_phenom}.  }
  \label{table_phenom0}
  \begin{tabular}{@{}lllll@{}}
  \hline
   Models                                &   Parameters                &      Obs 1          &         Obs 2       &   Obs 3            \\
 \hline
   Diffuse                               &    $N_{H}(1)$                      &    $1.42{\pm}0.03$      &   $1.25{\pm}0.03$ &  $0.58{\pm}0.01$     \\
   Plasma                                &    $(10^{22}$\,${\rm cm^{-2}})$    &                         &                   &                    \\
                                         &    ${\rm kT(1)}$\,(keV)            &    $={\rm Obs3}$        &   $={\rm Obs3}$   &  $0.96{\pm}0.02$ \\
                                         &    $N_{H}(2)$                      &    $0.13{\pm}0.02$      &   $0.18{\pm}0.02$ &  $0.06{\pm}0.02$   \\
                                         &    $(10^{22}$\,${\rm cm^{-2}})$    &                         &                   &                    \\
                                         &    ${\rm kT(2)}$\,(keV)            &    $={\rm Obs3}$        &   $={\rm Obs3}$   &  $0.60{\pm}0.02$   \\
 \hline
   Narrow Fe                             &    ${\rm E}_{\rm l}$\,(keV)        &    $={\rm Obs3}$        &    $={\rm Obs3}$  &  $6.7{\pm}0.3$     \\
   Line                                  &    ${\sigma}$\,(keV)               &    $={\rm Obs3}$        &    $={\rm Obs3}$  &  $0.06{\pm}0.04$    \\
 \hline
   Highecut                              &     ${\Gamma}$                     &    $1.47{\pm}0.03$      &   $1.49{\pm}0.03$ &  $1.79{\pm}0.02$    \\
   Powerlaw                              &     ${\rm E}_{\rm c}$\,(keV)       &    $-$                  &    $-$            &  $6.83{\pm}0.19$   \\
                                         &     ${\rm E}_{\rm f}$\,(keV)       &    $-$                  &    $-$            &  $9.7{\pm}1.4$     \\
  \hline
\end{tabular}
\end{table}

\section{The determination of the best model for the continuum intrinsic to the source}  \label{spec_anal2}

Considering the results obtained in the previous Section regarding the X-ray emission from the diffuse emission
of the galaxy, we are now in good condition to determine the best model for the intrinsic emission of the source.
We can see in Figure \ref{plots_phenom} and in the power-law photon indices obtained (see Table \ref{table_phenom0}) 
that the intrinsic spectrum of \m is variable. Thus we fitted each spectrum separately and we 
froze the parameters regarding the emission from the diffuse component from the model of the previous Section (i.e. 2 mekal components
plus narrow Fe line), expected to be constant. To this base-line model we added further models to describe the 
high-energy emission, as explained in the following.

The high-energy spectra are strongly curved, and we find that a model consisting by an absorbed power-law is not a good
description of the data of Obs. 3 (${\chi}^{2}/{\nu}=1.25,1.28,2.16$ with ${\nu}=1340,1399,730$ for Obs. 1, 2 and 3, respectively). The
results from the {\it Suzaku} spectrum are in agreement with the analysis from \citet{miyawaki09}. We considered X-ray emission 
from a multicolor disc ({\tt diskbb} in XSPEC) and Comptonization ({\tt CompTT} in XPSEC). 

The spectra of \m\ are convex, representing very well the prototype of ULXs of this class \citep{makishima07}. Indeed, the spectra are well fitted with emission
from a rather hot inner disc (${\rm kT}_{\rm in}=2.7-3.6$\,keV; see Table \ref{table_phenom}), values in agreement with previous studies 
\citep{stroh03,miyawaki09}. Nevertheless, these values are
significantly hotter than those obtained in recent observation (${\rm kT}_{\rm in}=0.9-1.6$) of \m\ by \citet{feng10}. This fact might correspond to the known
positive temperature-luminosity correlation known to act in ULXs \citep{makishima00}. But we do not observe a true change in the luminosities of \m\ with
respect those from \citet{feng10}, being very similar to the ones reported in this paper. \citet{feng10} identify their spectra as being from \m\ in a state
similar to the canonical high-soft state in BHBs (i.e. they follow the relationship ${\rm L}{\propto}{\rm T}^{4}$). \m\ might indeed be in a different state
during these observations, instead. \citet{mizuno07} reported the cut-off in the spectrum of NGC~1313 X--1 in the power-law state of ULXs and \citet{miyawaki09}
argued that \m\ was in the power-law state during Obs. 3, as well. Since there is barely any spectral evolution during these observations, we conclude that the
source was in the power-law state during all the observations reported in this paper.

The X-ray spectra of \m\ are reproduced very well by the {\tt compTT} model (see results of spectral fits in Table \ref{table_phenom}), 
invoking a rather low electron temperature
(${\rm kT}_{\rm e}=2.0-2.3$\,keV) and a large optical depth (${\tau}=7-23$). The results of Obs. 3 agree with those obtained previously for \m. 
The statistics reported in this work for Obs. 3 is poor (${\chi}^{2}/{\nu}{\ge}1.2$), since we considered the full 0.5--20\,keV energy range
in the fits of the {\it Suzaku} spectrum and below 2\,keV there are unavoidable residuals most probably due to calibration effects (and/or) the limitation
of our model to describe the diffuse emission from the galaxy. Since our goal is the study of the high-energy emission intrinsic to the source
we will just consider the 2--20\,keV energy range in the fits reported hereafter. 

We can see from the results of fitting with both X-ray curved models that the description with a Comptonization model is a better description just for
the spectrum of Obs. 1 (significant improvement by $8{\sigma}$). We thus consider the Comptonization model a better description of
the spectrum of Obs. 1. If luminosities from \m\ are truly super-Eddington, then the mentioned parameters suggest 
that the cold-corona model proposed by \citet{gladstone09} is a possibility in the description of the spectra of ULXs, emitting in the so-called Very-High state.
Nevertheless, this statement is just tentative and it is not required for the full data set analyzed in this paper. We will then consider both 
curved models ({\tt diskbb} and {\tt compTT}) as possible descriptions of the data and we will not discuss their validity any further.

\begin{table}
 \centering
  \caption{Models applied for the continuum intrinsic to \m (i.e. without contribution from the galaxy). Intrinsic luminosities are calculated in the 0.5--10\,keV and 2--20\,keV energy range for {\it XMM-Newton} and {\it Suzaku} spectra, respectively.  }
  \label{table_phenom}
  \begin{tabular}{@{}lllll@{}}
  \hline
   Models                                &   Parameters                &      Obs 1          &         Obs 2       &   Obs 3            \\
 \hline
   Diskbb                                &     ${\rm kT}_{\rm in}$\,(keV)  & $3.49{\pm}0.16$    &  $3.05{\pm}0.07$ & $2.73{\pm}0.06$                   \\
                                         &     ${\chi}^{2}/{\nu}$          & $1.19$             &  $1.17$          & $0.82$               \\
                                         &                                 & $(1589/1340)$      &  $(1636/1399)$   & $(465/566)$   \\
                                         &   $L_{\rm X}$\,(${\times}10^{40}$)       &  $3.22{\pm}0.02$   &  $2.05{\pm}0.20$ &  $4.8{\pm}0.4$       \\
                                         &   $({\rm erg\,s^{-1}})$     &                                     &                                   &                                       \\

  \hline
   CompTT                                &    ${\rm kT}_{\rm e}$\,(keV)&  $2.08{\pm}0.02$  &  $2.21{\pm}0.02$  &  $2.26{\pm}0.02$                \\
                                         &     ${\tau}$                &  $22.0{\pm}1.1$   &  $10.1{\pm}0.2$   &  $8.1{\pm}0.3$                 \\
                                         &     ${\chi}^{2}/{\nu}$      &  $1.05$           &  $1.14$           &  $0.82$            \\
                                         &                             &  $(1400/1338)$    &  $(1589/1397)$    &  $(465/564)$            \\
                                         & $L_{\rm X}$\,(${\times}10^{40}$)     &  $6.6{\pm}3.0$   &  $4.15{\pm}2.0$ &  $4.7{\pm}0.4$       \\
                                         &  $({\rm erg\,s^{-1}})$      &                                   &                                  &                                       \\
  \hline
\end{tabular}
\end{table}

\begin{figure}
\centering
 \includegraphics[bb=32 18 553 744,width=6cm,angle=270,clip]{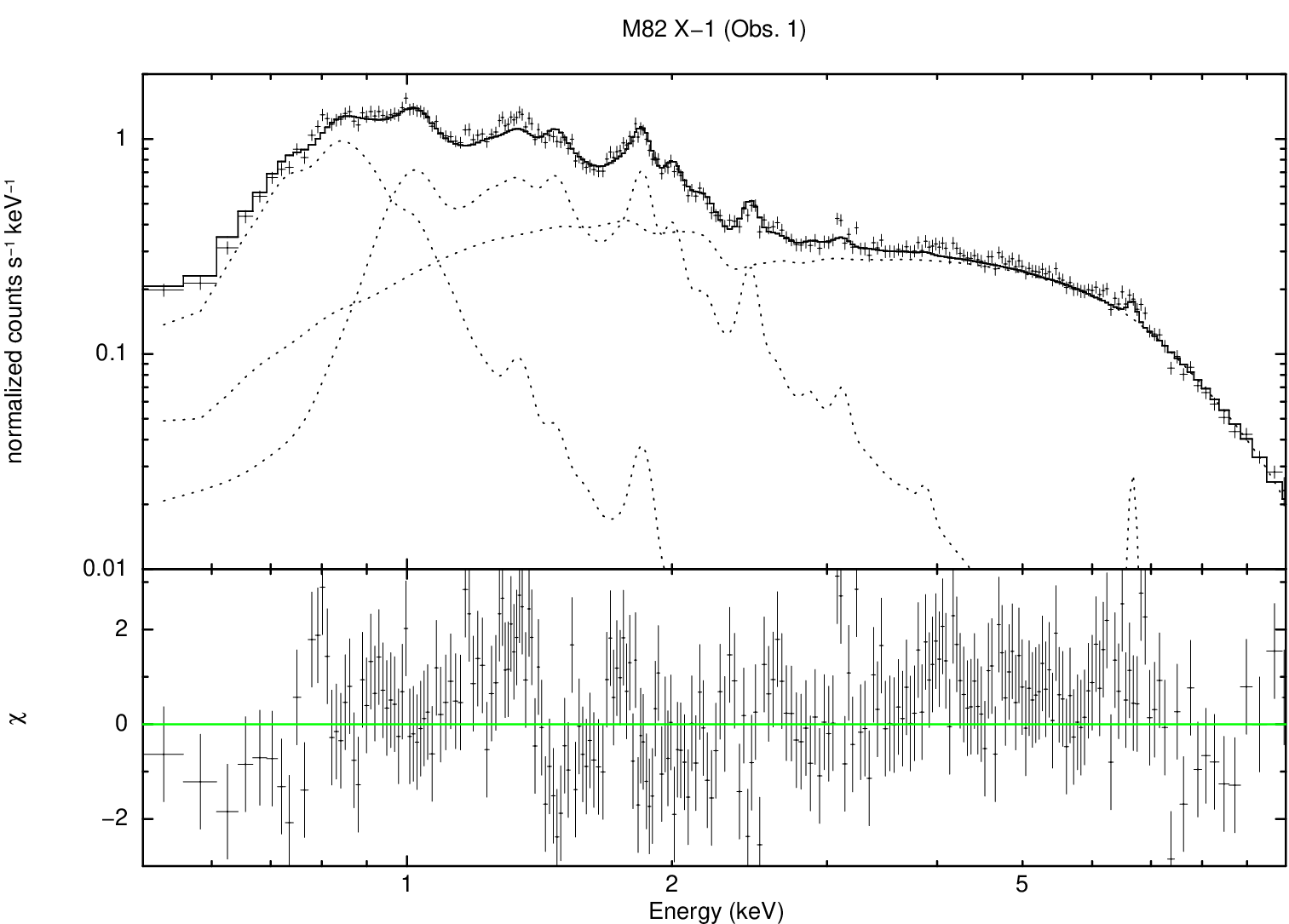}
 \includegraphics[bb=32 18 553 744,width=6cm,angle=270,clip]{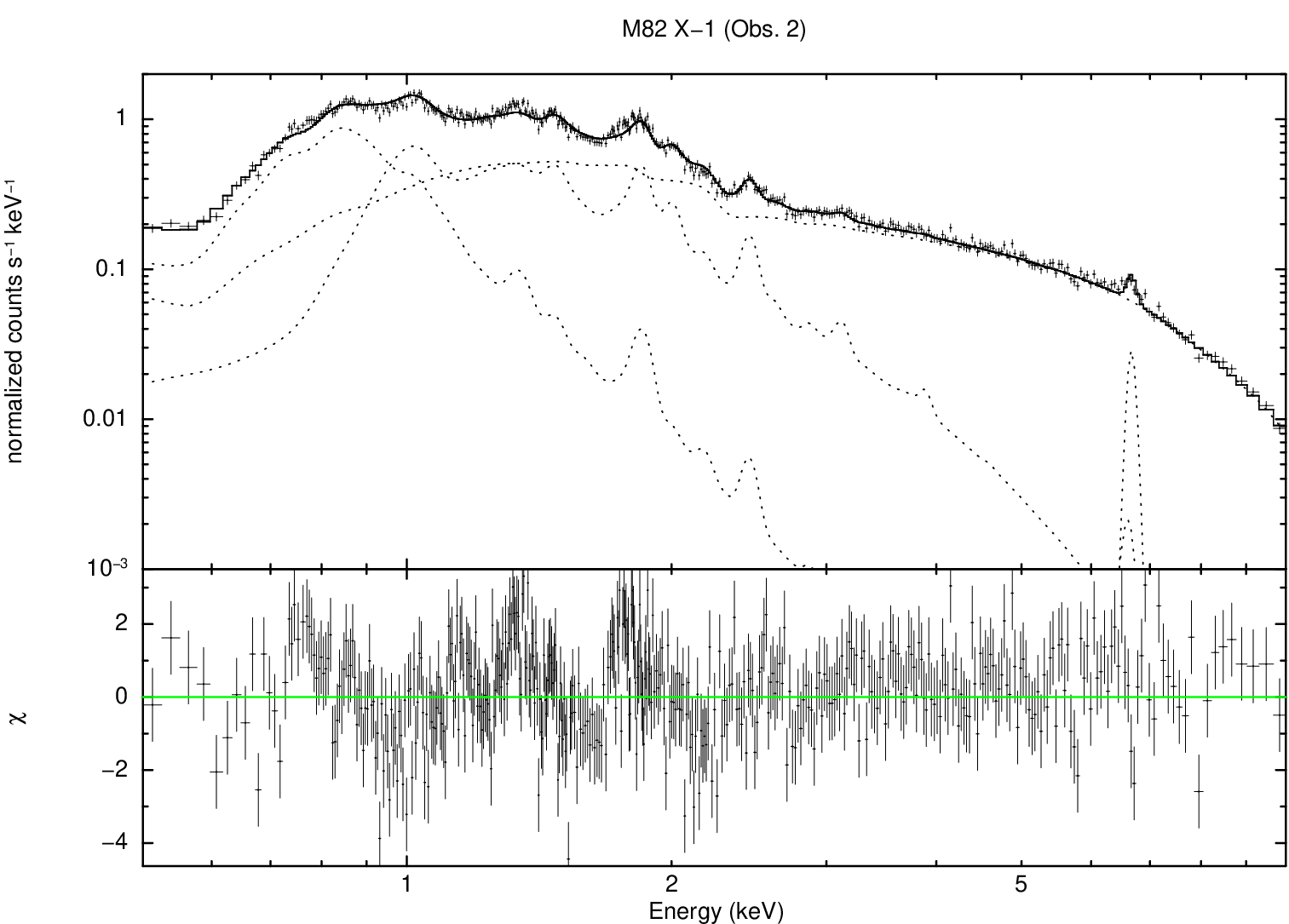}
 \includegraphics[bb=32 18 553 744,width=6cm,angle=270,clip]{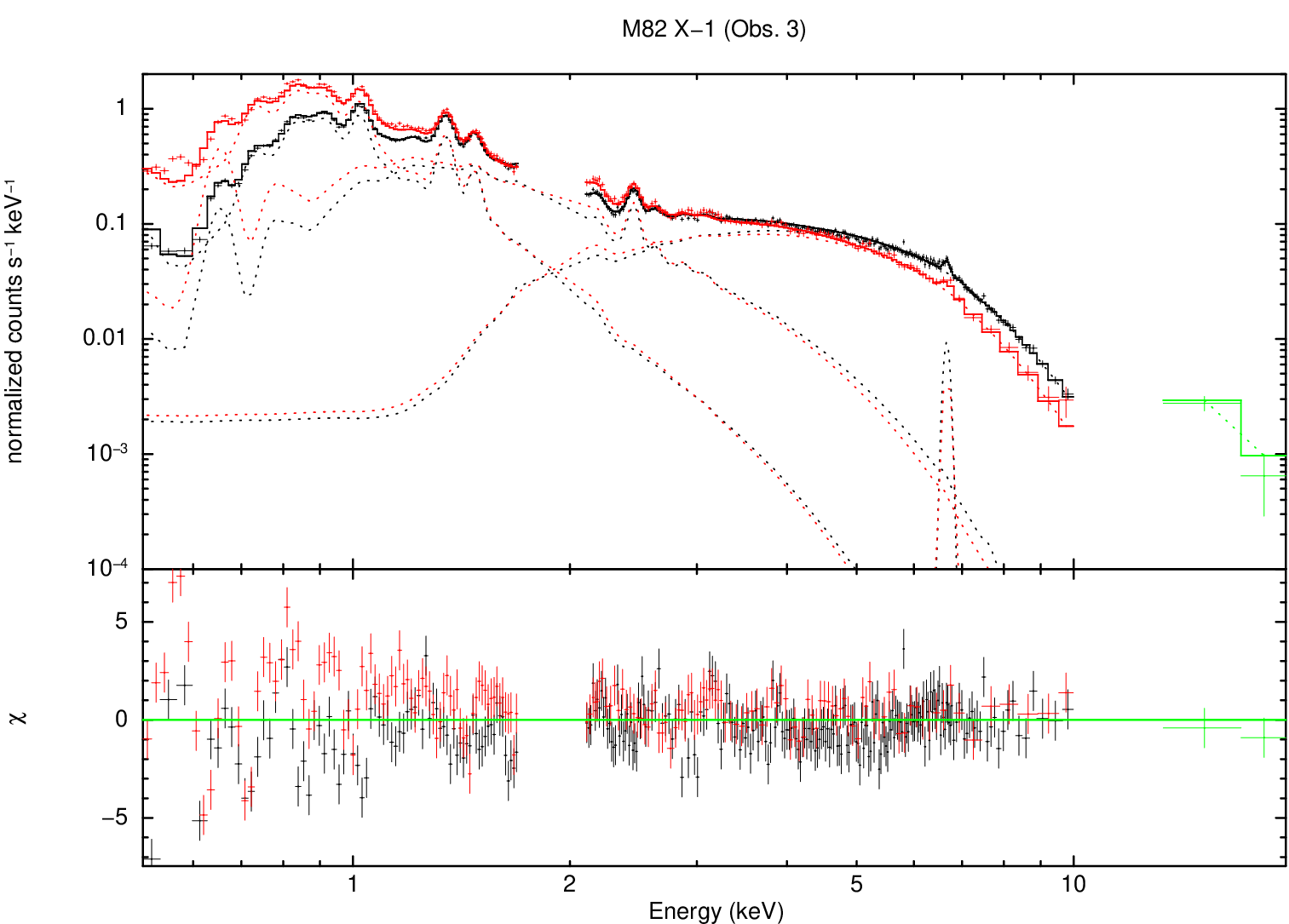}
 \caption{{\it XMM-Newton} and {\it Suzaku} spectra of \m\ (upper to lower) fitted with {\tt compTT} model. Continuous curvature of the overall emission and diffuse emission from two hot plasmas (${\rm kT}=0.6$\,keV and ${\rm kT}=0.96$\,keV) and galactic emission in the form of a narrow Fe line have been taken into account.}
 \label{plots_comptt}
\end{figure}

\section{Constraints of an emission broad Fe K${\alpha}$ line in the {\it Suzaku} spectrum}  \label{spec_anal3}

\begin{figure*}
\centering
 \includegraphics[bb=28 48 559 765,width=6cm,angle=270,clip]{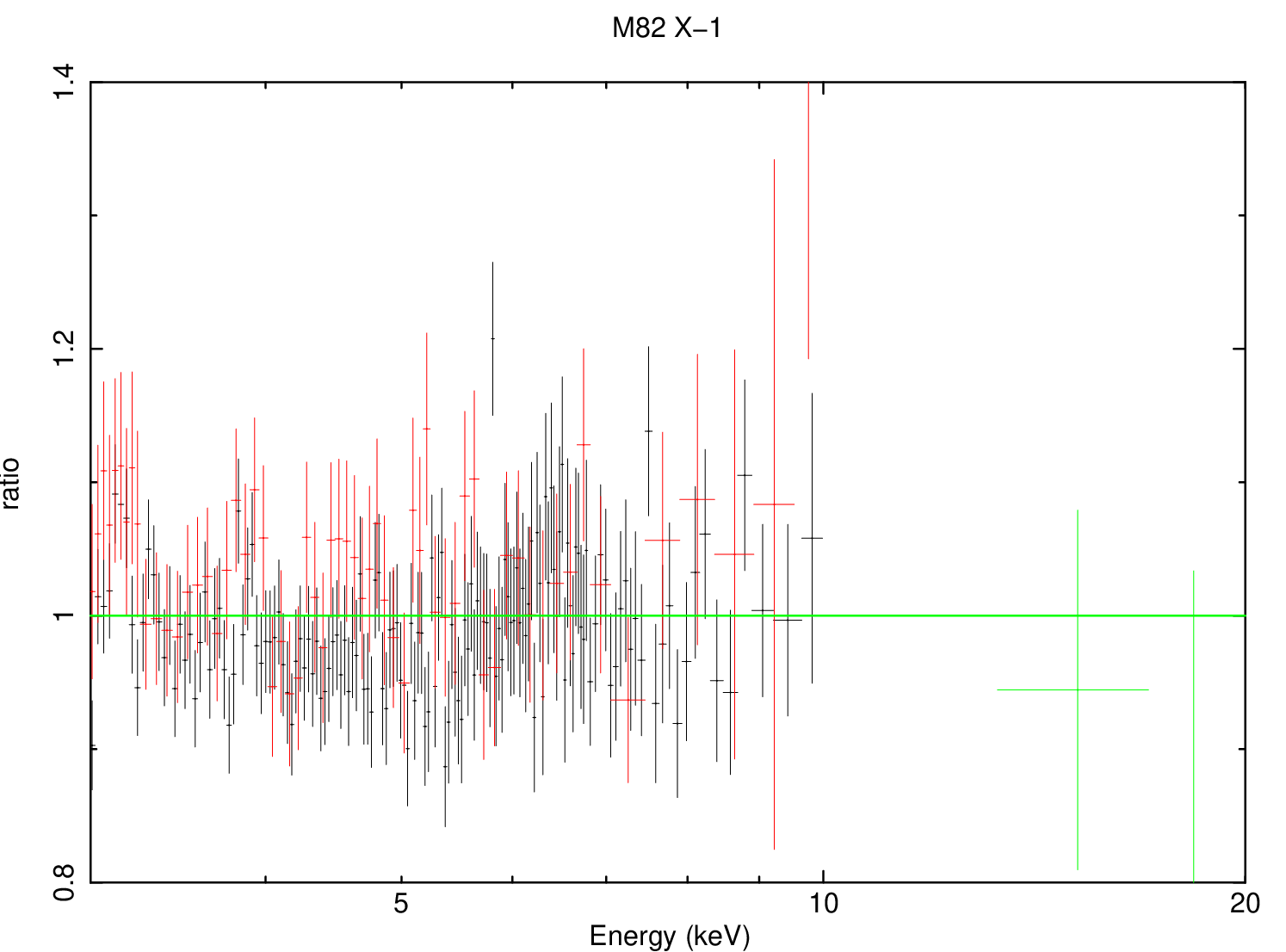}
 \includegraphics[bb=34 18 553 744,width=6cm,angle=270,clip]{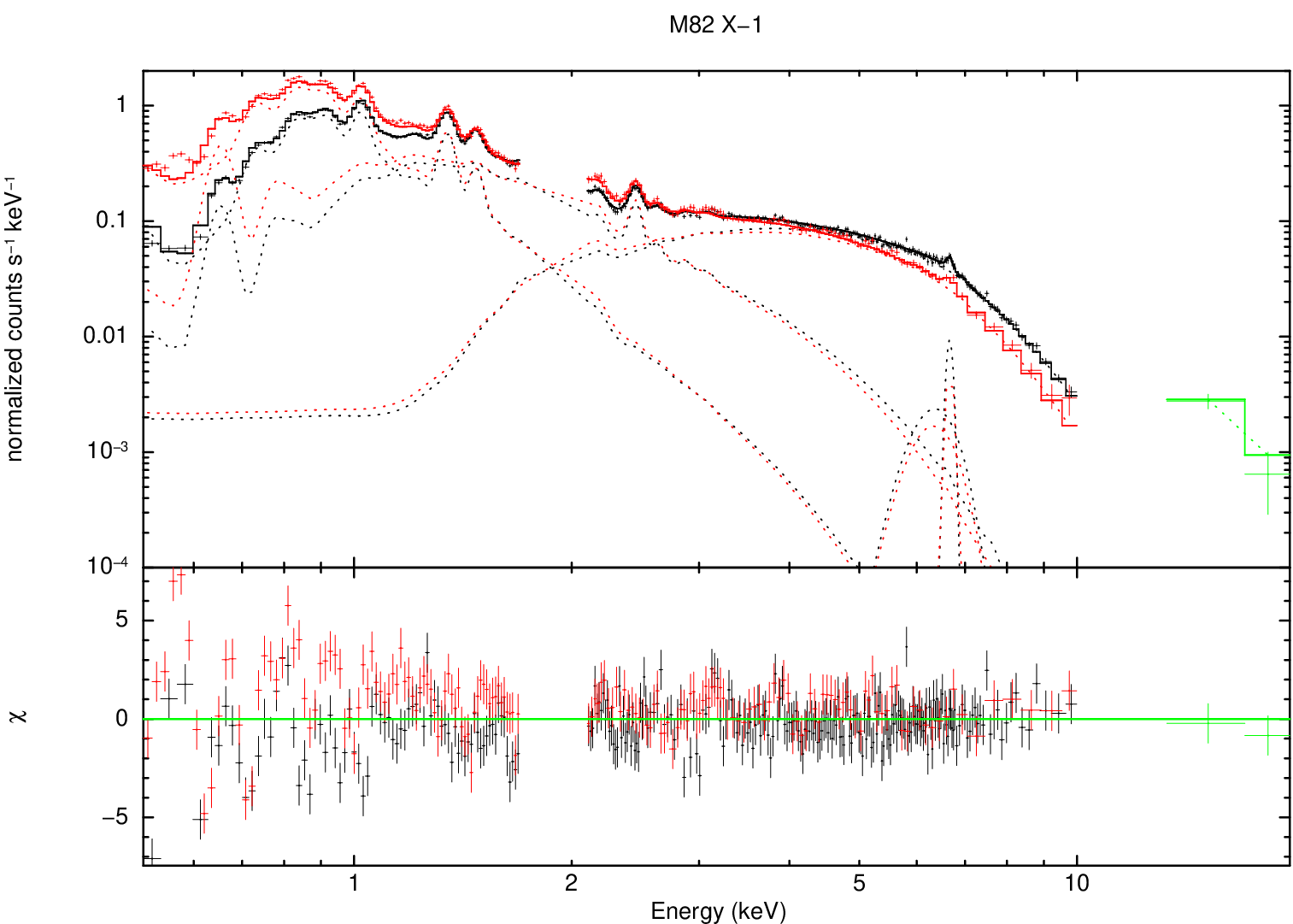}
 \caption{Ratio of the {\it Suzaku} spectrum with respect to the best model applied for the continuum (see text for details) (left) and spectrum with best fit model for the continuum plus the broad gaussian line component (right). }
 \label{plots_laor2}
\end{figure*}

The results of the previous Section with feasible models for the continuum leave some positive residuals in the 6--7\,keV energy range in
the {\it Suzaku}/XIS spectrum of Obs. 3 (see left panel of Figure \ref{plots_laor2}). These residuals could not be accounted for by the addition 
of the narrow Fe line and indicate the likely presence of a broad Fe emission feature. To analyze this possibility further, we added a broad
gaussian line to the best fit model (described in the previous Sections) of Obs. 3.  

The addition of a gaussian centred at 6.4--6.97\,keV in the best model describing the spectrum of Obs. 3
improves significantly the quality of the fit with any of the models used to describe the hard (${\ge}2$\,kev) X-ray spectra (improvement of 
${\Delta}{\chi}^{2}=13,22$ for ${\nu}=3$ additional degrees of freedom for Disc and Comptonization models, respectively). This means
a $3.3-4.4{\sigma}$ detection, which means a F-test probability of $<0.1\%$ that the improvement occurs by chance with any of both models.

The line is broad (${\sigma}=0.3-0.4$\,keV) and the derived equivalent width (EW) of the line is of ${\rm EW}=58^{+18}_{-30},59^{+25}_{-27}$\,eV
for the Disc and the Comptonization models, respectively. We conclude that the value of the equivalent width of the line is significantly
different from zero and that it is model-independent. Thus, we can consider the detection of the broad Fe line as a robust one, which does 
not depend upon the model considered for the continuum. In the right panel of Figure \ref{plots_laor2} we plot both the broad Fe line and the 
best description used for the underlying continuum. 

\subsection{Constraints on the spin of the Black Hole}

As shown above the presence of a broad Fe line is significant in the {\it Suzaku} spectrum (Obs. 3). The Equivalent-Width is in the range 30--80\,eV, 
in agreement with findings for some Active Galactic Nuclei, and it does not depend on the model applied for the continuum (Disc emission or 
Comptonization models). Accretion models predict that fluorescence
lines broadened by relativistic effects could arise from reflection of X-ray emission onto the inner region of the accretion
disc surrounding the black hole. In order to derive the inner disc radius and inclination we fitted the spectrum of Obs. 3 with a relativistic disc line 
profile calculated in a maximally spinning Kerr space-time ({\tt laor} model in XSPEC; \citealt{laor91}) but with the inner radius allowed to vary.
We froze the emissivity index of the power-law emission and the outer radius to $q=3$ and ${\rm R}_{\rm out}=100$\,${\rm R}_{\rm g}$, thus canonical 
values in BH-accreting systems. The inner disc radius was found to be ${\rm R}_{\rm in}=11_{-4}^{+9},10_{-4}^{+6}$\,${\rm R}_{\rm g}$ 
(values corresponding to the Disc and Comptonization models, respectively, for the continuum). 
This value is comparable to or even larger than the innermost stable circular orbit of a 
non-rotating (Schwarzschild) BH (i.e. $6\,{\rm R}_{\rm g}$). The inner disc inclination was found to be very small, too (${\rm i}=15_{-15}^{+9},10_{-10}^{+13}$\,$^{\circ}$), 
thus implying a close face-on configuration. Adding this line supposed an improvement in the fit of 
${\Delta}{\chi}^{2}=15,21$ for ${\nu}=4$ additional degrees of freedom for Disc and Comptonization models, respectively, thus very similar to the case of the
inclusion of the broad gaussian.

\section{Discussion and Conclusions}   \label{discuss}

In the present work we have studied the best broad-band X-ray spectra of \m currently available in X-rays from {\it XMM-Newton} and {\it Suzaku} satellites. 
With these high-quality data we have been able to disentangle emission from the diffuse plasma of the galaxy and intrinsic emission from 
the ULX. Our results imply that the X-ray diffuse emission is composed by a two-temperature thermal emission from a hot
plasma in multi-phase state. Similar temperatures of the components ($kT=0.96{\pm}0.02,0.60{\pm}0.02$\,keV) have been found for the highest 
S/N {\it ASCA} spectra of nearby elliptical galaxies \citep{buote98}. The intrinsic spectrum of the source obtained with {\it Suzaku} is 
significantly curved and properly accounted by models with intrinsic curvature (either emission from the inner accretion disc or Comptonization
from cold electrons in an optically thick corona), in agreement with previous studies. But the most important result of this paper is that
the (very) high quality observation with {\it Suzaku} has allowed us to detect a significant broad Fe K${\alpha}$ line in the spectrum of this 
notorious ULX. Assuming that the line feature is solely coming from the ULX via disc fluorescence, and that the inner radius reaches the last stable 
orbit, the data rules out a rapidly spinning BH and favours a non-spinning Schwarzschild-BH. Previous studies (\citealt{feng10}) have suggested
that the BH in this ULX is rapidly rotating but the possibility of a non-spinning BH has not been rejected so far.

The obtained equivalent width (EW) is in the range 30--80\,eV. This value for the EW (in Obs. 3) is notoriously lower than the tentative value 
previously reported (EW=0.26--0.43\,keV; \citealt{stroh03}) for the spectrum of Obs. 1. With the application of our model for the 
diffuse emission and the continuum of the source, we do not detect any Fe line in the same dataset of their work (Obs. 1). 
We have found that in the {\it Suzaku} spectrum this line is broad and the EW is similar to typical values found 
for the broad Fe K line in a wide sample of AGN \citep{delacalle10}. The value we found for the EW is significantly smaller
than those found in the typical cases of claimed relativistic lines (e.g., ${\approx}300$\,eV in MCG--6--30--15; \citealt{miniutti07}), for which an 
extremely fast rotating Kerr-BH has been claimed. Nevertheless, recent studies are setting much lower values \citep{noda11} for the spin of this AGN, 
underlying in the fact that the continuum was poorly modelled.

In the previous work of analysis of the {\it Suzaku} spectrum by \citet{miyawaki09} the broad Fe line was described by the addition of 
several narrow lines. This hypothesis looks reasonable, since a similar complex has been clearly seen in our Galactic Centre \citep{koyama07}. 
Nevertheless, the presence of relativistically broad Fe lines seems to be common in all types of accreting compact objects (neutron
stars, BHB, AGN) and evidence of their finding in ULXs would be very helpful in the understanding of the real physical processes accounting in these sources. 
Future longer X-ray observations will certainly provide further insights into the Fe complex and the presence of additional reflection features for 
this archetype of the ULX class.

\section*{Acknowledgments}
MCG is grateful to the anonymous referee and J. M. Miller, R. Soria and R. C. Reis for helpful 
comments and the provided Suzaku spectra, respectively. MCG acknowledges hospitality 
at {\it Departament de Astronomia i Meteorologia} ({\it Universitat de Barcelona},
Spain) and {\it Department of Physics} ({\it University of Crete}, Greece) during this work. 
This work is based on observations made with XMM-Newton, an
ESA science mission with instruments and contributions directly
funded by ESA member states and the USA (NASA). This research has made 
use of data obtained from the Suzaku satellite, a collaborative mission 
between the space agencies of Japan (JAXA) and the USA (NASA).

\end{document}